\documentclass{PoS}

\usepackage{amssymb,amsmath,amsfonts,amsthm}
\usepackage{array,multirow}
\usepackage{graphicx}
\usepackage{epstopdf} % This is for pdflatex

\newcommand{\Tr}{\operatorname{Tr}}

\renewcommand{\Re}{\operatorname{\mathfrak{Re}}}

\newcommand{\Fig}[1]{Fig.~\ref{#1}}
\newcommand{\Tab}[1]{Table~\ref{#1}}

\newcommand{\Eq}[1]{Eq.\,\eqref{#1}}

\graphicspath{{./figures/}}

\title{Coulomb gauge gluon propagator\\ on anisotropic lattices}

\ShortTitle{Coulomb gauge gluon propagator on anisotropic lattices}

\author{\speaker{Yoshiyuki Nakagawa}\\
        Research Institute for Information Science and Education,
        Hiroshima University, Higashi-Hiroshima, Hiroshima, 739-8521, Japan \\
        E-mail: \email{nkgw@rcnp.osaka-u.ac.jp}}

\author{Atsushi Nakamura\\
        Research Institute for Information Science and Education,
        Hiroshima University, Higashi-Hiroshima, Hiroshima, 739-8521, Japan \\
        E-mail: \email{nakamura@riise.hiroshima-u.ac.jp}}

\author{Takuya Saito\\
        Integrated Information Center, Kochi University\\
        Kochi, 780-8520, Japan\\
        E-mail: \email{tsaitou@kochi-u.ac.jp}}

\author{Hiroshi Toki\\
        Research Center for Nuclear Physics, Osaka University\\
        Ibarakisi, Osaka 567-0044, Japan\\
        E-mail: \email{toki@rcnp.osaka-u.ac.jp}}

\abstract{
We calculate the transverse and the time-time components
of the Coulomb-gauge gluon propagator
in SU(3) lattice Yang-Mills theory both on isotropic
and anisotropic lattices.
The problem of scaling violation observed on the isotropic lattice
is drastically reduced as the anisotropy increases;
namely, the system approaches the Hamiltonian limit.
In the infrared region, the transverse gluon propagator exhibits
a turnover and the temporal gluon propagator shows divergent behavior.
          }

\FullConference{The XXVII International Symposium on Lattice Field Theory - LAT2009\\
		 July 26-31 2009\\
		 Peking University, Beijing, China}

\begin{document}

%=======================================================
\section{Introduction}
\label{sec:introduction}
%=======================================================

Coulomb gauge provides a very clear picture of color confinement.
Coulomb gauge is a physical gauge in the sense that
the color Gauss' law can be formally solved,
and only transverse degrees of freedom appears as
dynamical degrees of freedom.
The striking feature of Coulomb gauge is that
an instantaneous interaction shows up in the Hamiltonian,
which is requisite for color confinement.
In the Gribov-Zwanziger scenario, the path integral is dominated
by the configurations near the Gribov horizon where the lowest
eigenvalue of the Faddeev-Popov (FP) ghost operator vanishes.
\cite{ZwanzigerD:NPB412:1994}.
This results in an enhancement of the near-zero modes of
the ghost operator, and it has been confirmed by the lattice simulations
\cite{GreensiteJ:JHEP05:2005,NakagawaY:PRD75:2007}.
Accordingly, the color-Coulomb instantaneous interaction
becomes a confining interaction.
The color-Coulomb potential can be obtained by measuring
the correlator of the temporal link variable at fixed time,
and lattice QCD simulations exhibit that the color-Coulomb potential
rises linearly at large distances and its string tension is larger
than the string tension of the static Wilson potential
\cite{GreensiteJ:PRD67:2003,NakamuraA:PTP115:2006,
Nakagawa:2006fk,NakagawaY:PRD77:2008},
which is expected from the Zwanziger's inequality
\cite{ZwanzigerD:PRL90:2003}.
On the other hand, the color-Coulomb potential has been
evaluated by inverting the FP ghost matrix, and it has been shown that
the color-Coulomb string tension almost saturates
the Wilson string tension
\cite{VoigtA:PRD78:2008}.

The transverse gluon propagator is expected to be suppressed
in the infrared (IR) region due to the proximity of the Gribov region
in the IR direction in the Gribov-Zwanziger scenario
\cite{ZwanzigerD:NPB364:1991}.
The instantaneous transverse gluon propagator
has been measured by Monte Carlo simulations
\cite{CucchieriA:PRD65:2001,LangfeldK:PRD70:2004,
BurgioG:PRL102:2009,NakagawaY:PRD79:2009},
and recent studies have revealed that it shows scaling violation
\cite{BurgioG:PRL102:2009,NakagawaY:PRD79:2009};
namely, the gluon propagator calculated at different lattice couplings
does not fall on top of a single curve after multiplicative renormalization.

In order to circumvent the problem of scaling violation,
the authors of \cite{BurgioG:PRL102:2009}
have measured the unequal-time gluon propagator
\begin{equation}
D^{\textrm{tr}}(\vec{p},p_4) = \langle A(\vec{p},p_4) A(-\vec{p},-p_4) \rangle
\end{equation}
and extracted the equal-time propagator $D^{\textrm{tr}}(|\vec{p}|)$
by eliminating the $p_4$ dependence of the unequal-time propagator.
It has been concluded that
$D^{\textrm{tr}}(|\vec{p}|)$ is multiplicatively renormalizable
in the Hamiltonian limit and well fitted with the Gribov-type form
of the propagator.
The method was applied only to the transverse gluon propagator
and it was not studied if scaling violation can be solved
by this procedure.

In \cite{NakagawaY:PRD79:2009},
a new momentum cut is introduced in addition to the cone cut
and the cylinder cut, by which high momentum data that suffer
from discretization errors are excluded
from the analysis of the instantaneous propagators.
It has been shown that this procedure successfully reduces
scaling violation for the transverse gluon propagator while
it fails for the time-time component of the gluon propagator.

The problem of scaling violation of the instantaneous propagator
can be seen even at the tree level on a finite temporal lattice spacing
(we refer to a forthcoming paper for an explicit calculation).
The reason is that the energy integral does not run from $-\infty$
to $\infty$ but from $-2/a_{\tau}$ to $2/a_{\tau}$ on a finite lattice,
and this introduces the spurious $|\vec{p}|$ dependence on
the free equal-time propagator.
Therefore, we expect that the instantaneous propagator is
multiplicatively renormalizable in the Hamiltonian limit
$\xi=a_s/a_{\tau}\to\infty$.
To make this point clear, we calculate the transverse and temporal
components of the instantaneous gluon propagator on anisotropic lattices.

%=======================================================
\section{Lattice setup and observables}
\label{sec:lattice_setup}
%=======================================================

The lattice configurations are generated by
the heat-bath Monte Carlo technique with
the standard Wilson plaquette action,
\begin{equation}
  S = \frac{\beta}{\xi_B} \sum_{n, i < j \le 3}
  \Re\Tr (1-U_{ij}(n)) \\
    + \beta \xi_B \sum_{n, i \le 3}
  \Re\Tr (1-U_{i4}(n)).
\end{equation}
Here $U_{\mu\nu}(n)$ indicates the plaquette operator, and
$\beta=2 N_c / g^2$ is the lattice coupling.
On the isotropic lattice, the bare anisotropy $\xi_B$ is 1
and the action can be written in a familiar form
\begin{equation}
  S = \beta \sum_{n, \mu < \nu} \Re\Tr (1-U_{\mu\nu}(n)). \\
\end{equation}
$\xi_B$ differs from the renormalized anisotropy $\xi$ which is
defined as the ratio of the spatial lattice spacing to
the temporal lattice spacing.
The ratio of $\xi_B$ and $\xi$ can be determined non-perturbatively by
matching the spatial and the temporal Wilson loop on anisotropic lattices.
We use the relation obtained by Klassen for the range $1 \le \xi_B \le 6$
and $5.5 \le \beta \le \infty$
\cite{KlassenTR:NPB533:1998}:
\begin{equation}
  \frac{\xi}{\xi_B} = 1 + \left( 1 - \frac{1}{\xi} \right)
  \frac{\eta (\xi)}{6} \frac{1 + a_1 g^2}{1 + a_0 g^2} g^2,
\end{equation}
where $a_0=-0.77810$, $a_1=-0.55055$, and
\begin{equation}
  \eta(\xi) = \frac{1.002503 \xi_B^3 + 0.39100 \xi_B^2 + 1.47130 \xi_B - 0.19231}
                            {\xi_B^3 + 0.26287 \xi_B^2 + 1.59008 \xi_B - 0.18224}.
\end{equation}
We adopt the values of the lattice spacing given in
\cite{NamekawaY:PRD64:2001}
for $\xi=2$ and in
\cite{MatsufuruH:PRD64:2001}
for $\xi=4$,
where the static quark potential was measured
to set the scale.
For the isotropic lattice, the scale is set by using
Necco-Sommer scaling relation
\cite{NeccoS:NPB622:2002}.
In our simulations,
the first 5000 sweeps are discarded for thermalization,
and we measured the equal-time gluon propagator for 100 configurations,
each of which is separated by 100 sweeps.
All the lattice parameters are given in \Tab{tab:lattice_setup}.

%**************************************************%
\begin{table}[htbp]
\begin{center}\begin{tabular}{cccccccc}
\hline\hline 
$\xi=a_s/a_{\tau}$ & $L_s^3 \times L_{\tau}$ & $\beta$ & $\xi_B$
& $a_s^{-1}$ [GeV] & $a_s$ [fm] & $V$[fm$^4$] & \# of confs. \\
\hline \hline
\multirow{6}{*}{1}
& 32$^4$           & 5.70 & 1     & 1.160 & 0.1702 & 5.45$^4$ & 100 \\
& 48$^4$           &  :   &   :   &   :   &   :    & 8.17$^4$ &  40 \\
& 32$^4$           & 5.80 &   :   & 1.446 & 0.1364 & 4.37$^4$ & 100 \\
& 32$^4$           & 6.00 &   :   & 2.118 & 0.0932 & 2.98$^4$ & 100 \\
& 48$^4$           &  :   &   :   &   :   &   :    & 4.47$^4$ &  40 \\
& 32$^4$           & 6.20 &   :   & 2.914 & 0.0677 & 2.17$^4$ & 100 \\
\hline
\multirow{6}{*}{2}
& 16$^3\times$ 32  & 5.80 & 1.674 & 1.104 & 0.1787 & 2.86$^4$ & 100 \\
& 24$^3\times$ 48  &  :   &   :   &   :   &    :   & 4.29$^4$ & 100 \\
& 16$^3\times$ 32  & 6.00 & 1.705 & 1.609 & 0.1227 & 1.96$^4$ & 100 \\
& 24$^3\times$ 48  &  :   &   :   &   :   &    :   & 2.94$^4$ & 100 \\
& 16$^3\times$ 32  & 6.10 & 1.718 & 1.889 & 0.1045 & 1.67$^4$ & 100 \\
& 24$^3\times$ 48  &  :   &   :   &   :   &    :   & 2.51$^4$ & 100 \\
\hline
\multirow{12}{*}{4}
& 16$^3\times$ 64  & 5.75 & 3.072 & 1.100 & 0.1794 & 2.87$^4$ & 50 \\
& 24$^3\times$ 96  &  :   &   :   &   :   &    :   & 4.31$^4$ & 50 \\
& 32$^3\times$ 128 &  :   &   :   &   :   &    :   & 5.74$^4$ & 50 \\
& 48$^3\times$ 192 &  :   &   :   &   :   &    :   & 8.61$^4$ & 50 \\
& 16$^3\times$ 64  & 5.95 & 3.159 & 1.623 & 0.1216 & 1.95$^4$ & 50 \\
& 24$^3\times$ 96  &  :   &   :   &   :   &    :   & 2.92$^4$ & 50 \\
& 32$^3\times$ 128 &  :   &   :   &   :   &    :   & 3.89$^4$ & 50 \\
& 48$^3\times$ 192 &  :   &   :   &   :   &    :   & 5.84$^4$ & 50 \\
& 16$^3\times$ 64  & 6.10 & 3.211 & 2.030 & 0.0972 & 1.56$^4$ & 50 \\
& 24$^3\times$ 96  &  :   &   :   &   :   &    :   & 2.33$^4$ & 50 \\
& 32$^3\times$ 128 &  :   &   :   &   :   &    :   & 3.11$^4$ & 50 \\
& 48$^3\times$ 192 &  :   &   :   &   :   &    :   & 4.67$^4$ & 50 \\
\hline\hline
\end{tabular}
\caption{
Simulation parameters to calculate the equal-time gluon propagator.
}
\label{tab:lattice_setup}
\end{center}\end{table}
%**************************************************%

In Coulomb gauge the transversality condition
\begin{equation}
\partial_i A_i(\vec{x},t) = 0
\end{equation}
is imposed on the gauge fields at each time slice,
where $i$ runs from 1 to 3.
On a lattice, gauge configurations satisfying Coulomb gauge
condition can be obtained by minimizing the functional
\begin{equation}\label{F_U}
F_U[g] = \sum_{i=1}^3 \sum_{\vec{x}} \Re\Tr \left( 1 - U_i^g(\vec{x},t) \right),
\end{equation}
defined on each time slice.
Here
$U_i^g(\vec{x},t) = g(\vec{x},t) U_i(\vec{x},t) g^{\dagger}(\vec{x}+\hat{i},t)$
is the gauge-rotated configuration.
The functional derivative of \Eq{F_U} with respect to $g$
reproduces the Coulomb gauge condition in the continuum limit
with the linear definition of the gauge field,
\begin{equation}\label{eq:linear_def}
A^{\textrm{lat}}_{\mu}(\vec{x},t)
= \left.
\frac{U_{\mu}(\vec{x},t)-U_{\mu}^{\dagger}(\vec{x},t)}{2iga}
\right|_{\textrm{traceless}}. 
\end{equation}
The Coulomb gauge fixing has been done using iterative method
with the Fourier acceleration
\cite{Davies:1987vs},
and the gauge fixing is stopped if
$(\partial_iA_i)^2 < 10^{-14}$ at each time slice.

We calculate the transverse and time-time component
of the equal-time gluon propagator,
\begin{equation}
D^{ab}_{\mu\nu}(\vec{x}-\vec{y})
= \langle A^a_{\mu}(\vec{x})A^b_{\nu}(\vec{y})\rangle
= D^{ab}_{\mu\nu}(\vec{x}-\vec{y}),
\end{equation}
in the momentum space,
\begin{align}
& D^{ab}_{ij}(\vec{p})
  = \delta^{ab}\left(\delta_{ij}-\frac{p_ip_j}{|\vec{p}|^2}\right)
  D^{\mathrm{tr}}(|\vec{p}|) \\
& D^{ab}_{44}(\vec{p})
  =\delta^{ab}\frac{Z^{44}(|\vec{p}|)}{|\vec{p}|^2}.
\end{align}
The gauge field is defined as \Eq{eq:linear_def}.
The dressing function $Z^{44}$ is constant for $|\vec{p}|$ at the tree level.
In the Gribov-Zwanziger scenario, this is expected to diverge in
the IR limit resulting in the confining behavior of
the color-Coulomb potential, which is necessary condition for color confinement
in Coulomb gauge QCD.
The transverse gluon propagator is expected to be suppressed
in the IR region due to the proximity of the Gribov region
\cite{ZwanzigerD:NPB364:1991}.

%=======================================================
\section{Simulation results: transverse gluon propagator}
\label{sec:results_Dtr}
%=======================================================

%**************************************************%
\begin{figure}[htbp]
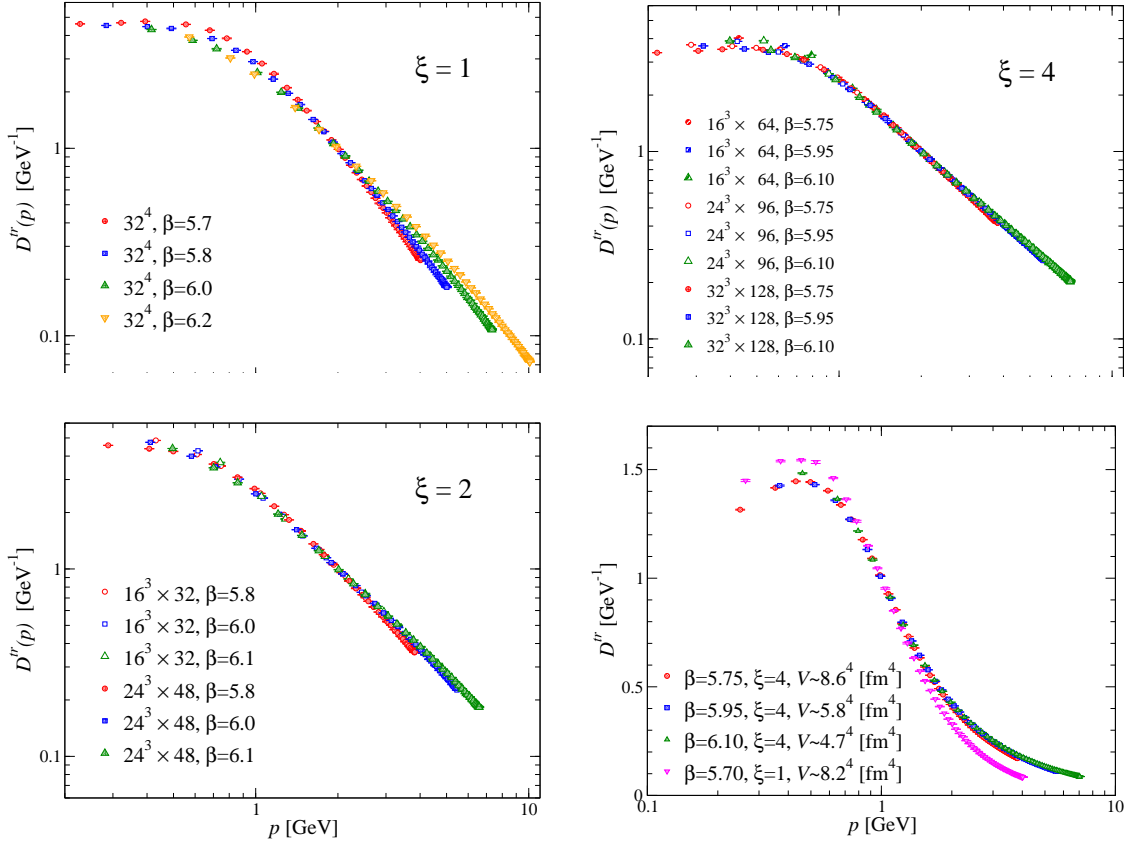

\begin{minipage}{0.47\hsize}\begin{center}
\resizebox{1.\textwidth}{!}
{\includegraphics{Dtr_rnmlzd_3232}}
\resizebox{1.\textwidth}{!}
{\includegraphics{Dtr_rnmlzd_aniso_xi2}}
\end{center}\end{minipage}
\hspace{0.03\hsize}
\begin{minipage}{0.47\hsize}\begin{center}
\resizebox{1.\textwidth}{!}
{\includegraphics{Dtr_rnmlzd_aniso_xi4}}
\resizebox{1.\textwidth}{!}
{\includegraphics{Dtr_renorm_aniso}}
\end{center}\end{minipage}
\caption{
The equal-time transverse gluon propagator
on the isotropic lattice (top left),
on the anisotropic lattices with $\xi=2$ (bottom left),
and with $\xi=4$ (top right).
The results for the isotropic lattice and the anisotropic lattice
with $\xi=4$ on large lattice volume are drawn together
in one figure for direct comparison
(bottom right).
The cone cut and the cylinder cut are applied and
the propagator is renormalized to unity at $p=2$ [GeV].
}
\label{Dtr_renormalized}
\end{figure}
%**************************************************%

The top left panel of \Fig{Dtr_renormalized} shows
the transverse gluon propagator
on the isotropic lattice at various lattice couplings,
$\beta=5.7, 5.8, 6.0, 6.2$.
The cone cut and the cylinder cut are applied
\cite{LeinweberDB:PRD60:1999}
and the propagator is renormalized such that
$D^{tr}(|\vec{p}|=2\textrm{ [GeV]}) = 1$.
We see that the data points at different
lattice couplings cross at the renormalization point
$|\vec{p}|=2$ [GeV] and deviate from each other both
at small and large momenta, as has been observed in 
\cite{NakagawaY:PRD79:2009}.

The simulation results on the anisotropic lattice are drawn
in the left bottom ($\xi=2$)
and the right top ($\xi=4$) panel of \Fig{Dtr_renormalized}.
On the anisotropic lattice with $\xi=2$,
scaling violation becomes moderate compared to the isotropic result,
although small deviations among the data points
for different lattice couplings can be seen.
Further increase of $\xi$ leads to a nice scaling behavior
and the data points for $\xi=4$ almost fall on top of one curve,
indicating that the equal-time gluon propagator is multiplicatively
renormalizable in the continuum limit 
(or in the Hamiltonian limit $\xi\to\infty$).
Accordingly, our results on the anisotropic lattice support
our expectation that scaling violation observed in the equal-time
transverse gluon propagator disappears in the limit $\xi\to\infty$.

In the right bottom panel of \Fig{Dtr_renormalized},
the instantaneous transverse gluon propagator
on the spatial lattice extent $L=48$ is plotted
both for the isotropic lattice and the anisotropic lattice with $\xi=4$.
We observe that the propagator has a maximum at $p = 0.4 \sim 0.5$ [GeV]
irrespective of the lattice coupling and the anisotropy,
and it decreases with the momentum in the IR region.

%=======================================================
\section{Simulation results: time-time component of the gluon propagator}
\label{sec:results_Z44}
%=======================================================

%**************************************************%
\begin{figure}[htbp]
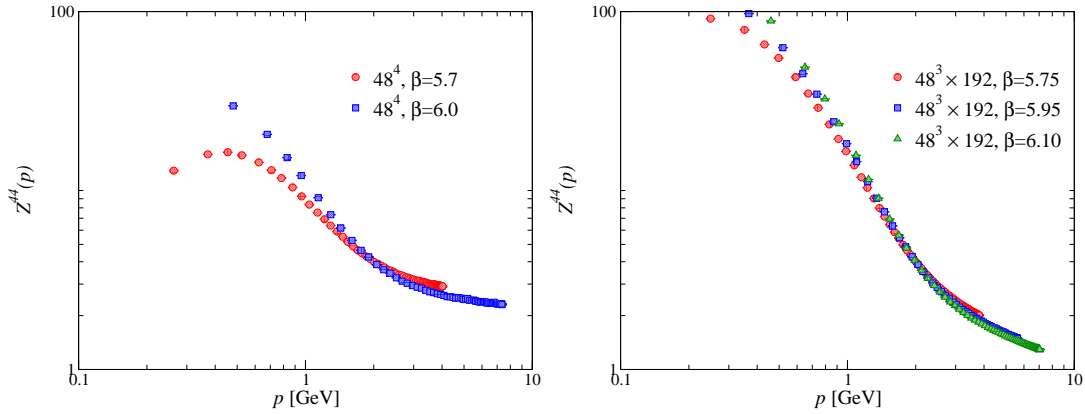

\begin{minipage}{0.47\hsize}\begin{center}
\resizebox{1.\textwidth}{!}
{\includegraphics{Z44_on48x48_iso}}
\end{center}\end{minipage}
\begin{minipage}{0.47\hsize}\begin{center}
\resizebox{1.\textwidth}{!}
{\includegraphics{Z44_on48x192_aniso}}
\end{center}\end{minipage}
\caption{
The dressing function of
the time-time component of the gluon propagator
on the isotropic lattice (left), and
on the anisotropic lattice with $\xi=4$ (right).
The cone cut and the cylinder cut are applied and
the dressing function is renormalized to unity at $p=2$ [GeV].
}
\label{Z44_renormalized}
\end{figure}
%**************************************************%

The dressing function of
the time-time component of the gluon propagator is shown
in \Fig{Z44_renormalized} for the isotropic lattice (left panel)
and the anisotropic lattice with $\xi=4$ (right panel).
On the isotropic lattice, $Z^{44}(|\vec{p}|)$ shows scaling violation
and the deviation of the two curve is pronounce in the IR region.
Although the Gribov-Zwanziger scenario predicts that the temporal
gluon propagator diverges stronger than the simple pole $1/|\vec{p}|^2$,
the numerical result shows that it bends down at small momenta for $\beta=5.7$.

On the anisotropic lattice, the dressing function shows
a much better scaling behavior than that on the isotropic lattice.
Although the small deviation can be seen both in the IR and
ultraviolet region, one can expect that the scaling behavior
is completely recovered in the Hamiltonian limit.
Moreover, we find that the IR behavior of $Z^{44}$
on the anisotropic lattice is completely different from that
on the isotropic lattice.
For the isotropic case, we see that the dressing function
bends down at small momenta at $\beta=5.7$. 
By contrast, $Z^{44}$ continues to rise with decreasing the momentum
even for the coarsest lattice data ($\beta=5.75$),
and $Z^{44}$ at available smallest momentum for the anisotropic case
is about 10 times larger than that for the isotropic case.
We note that the spatial lattice spacing for $(\xi,\beta)=(4,5.75)$ is
larger than that for $(\xi,\beta)=(1,5.70)$.
This implies that $Z^{44}$ is very sensitive to the disretization effects,
and taking the Hamiltonian limit is crucial to cure scaling violation
for the temporal gluon propagator and to explore the IR divergent behavior
in Coulomb gauge QCD.

%=======================================================
\section{Summary and conclusion}
\label{sec:Summary}
%=======================================================

We calculate the transverse and time-time components of
the equal-time gluon propagator
both on the isotropic and the anisotropic lattices.
We find that scaling violation observed on the isotropic lattice
is drastically reduced by calculating the propagator
on the anisotropic lattices,
i.e., by getting close to the Hamiltonian limit.
In the IR region, the transverse gluon propagator is strongly
suppressed and shows the turnover at about 500 [MeV].
The time-time gluon propagator on the anisotropic lattice
is much more enhanced in the IR region compared to
that on the isotropic lattice.

%=======================================================
\section*{Acknowledgements}
%=======================================================

The simulation was performed on
NEC SX-8R at RCNP, and NEC SX-9 at CMC, Osaka University.
We appreciate the warm hospitality and support of the RCNP administrators.
Y. N. is supported by Grant-in-Aid for JSPS Fellows
from Monbu-kagakusyo.

%%%%%%%%%%%%%%%%%%%%%%%%%%%%%%%%%%%%%%%%%%%%%%%%%%%%%%%%

%%%%%%%%%%%%%%%%%%%%%%%%%%%%%%%%%%%%%%%%%%%%%%%%%%%%%%%%

\end{document}